\documentstyle[twocolumn,aps]{revtex}
\newcommand{\half}{\frac{1}{2}}

\begin{document}
% \draft command makes pacs numbers print
%\draft
%\twocolumn
\title{ Nonlocal potentials for short-range electronic correlation
in atoms, molecules, and solids}
% repeat the \author\address pair as needed
\author{R. K. Nesbet}
%\affiliation{
\address{
Physics Department, 
University of Connecticut,
Storrs, CT 06269-3046, USA}
\address{Permanent address:
IBM Almaden Research Center,
650 Harry Road,
San Jose, CA 95120-6099, USA}
\date{\today}
%\widetext
\maketitle
\begin{center}For {\em Int.J.Quantum Chem.} \end{center}
\begin{abstract}
% insert abstract here
Extending density functional theory (DFT) to an {\it ab initio} orbital
functional theory (OFT) requires new methodology for nonlocal exchange
and correlation potentials.  This paper describes such modifications to 
a standard Dirac-Slater atomic program.  Unrestricted Hartree-Fock (UHF)
theory is extended by a modified Colle-Salvetti Ansatz for short-range 
electronic correlation.  Results are reported for atoms He-Ne. Values of
parameters needed for similar calculations on molecules and solids are
reported.  Implementation of nonlocal exchange and correlation for
such extended systems, using multiple scattering theory to connect 
independent calculations in space-filling atomic cells, is discussed.
\end{abstract}
\section{Introduction} 
Without invoking the electronic density as an intermediate, variational
equations for an independent-electron model can be derived as an
orbital functional theory (OFT)\cite{NES01}, in agreement with
Kohn-Sham equations in the local density approximation (LDA).  
In contrast to density functional theory (DFT)\cite{HAK64,KAS65}, 
an exact energy functional for OFT can be derived from 
many-body theory\cite{NES00,NES01a}.  The orbital Euler-Lagrange 
equations (OEL) of OFT are determined by orbital functional derivatives,
which take the general form of linear operators acting on occupied 
orbital functions of a model state.  Because the exclusion principle 
requires independent normalization of the orbital partial densities,
derivation of the OEL equations requires orbital wave functions or 
densities to be varied independently\cite{NES02}.  For N 
electrons, there are N normalization constraints. Independent Lagrange 
multipliers (the orbital energy eigenvalues) are determined by these 
normalization constraints, as they are for Hartree-Fock and 
Kohn-Sham equations.  
\par In a density-based theory that incorporates the exclusion 
principle\cite{NES02a}, Euler-Lagrange equations for the orbital 
densities are determined by partial (G\^ateaux\cite{BAB92}) density 
functional derivatives, indexed by the orbital densities.  Unless they 
are independent of the orbital index, these functional derivatives 
do not determine a unique total (Fr\'echet\cite{BAB92}) density 
functional derivative, equivalent to a multiplicative local potential 
function\cite{NES02,NES98}.  The required G\^ateaux functional 
derivatives can be constructed from the corresponding orbital functional
derivatives of OFT.  The implied generalized Thomas-Fermi equations 
for orbital partial densities are operationally equivalent to the OEL 
equations\cite{NES02a}.  
\par Unless all orbital energies are equal\cite{NES02,NES98}, the OEL 
equations cannot be derived by a theory  which considers only the total 
electronic density\cite{NES03a}.  Equal orbital energies conflict with 
the exclusion principle for the lowest state of any compact system with 
more than one electron of each spin.  This is easily verified in
the example of the $1s2s\,^3S$ state of an atom with two noninteracting 
electrons of the same spin.  Independent variation of orbital 
densities, required by the exclusion principle, defines G\^ateaux
functional derivatives.  It follows from this analysis that extension of
Kohn-Sham theory beyond the LDA cannot be done in general without 
introducing nonlocal potentials\cite{NES03a}.  The required nonlocal 
potentials are well-defined in OFT, formulated as an application of the 
many-body theory of electrons.  
\par The Schr\"odinger kinetic energy operator ${\hat t}=-\half\nabla^2$
and the Fock exchange operator are well-known nonlocal "potentials" in
standard theory.  The present work includes nonlocal potentials 
for short-range correlation based on the correlation energy Ansatz of 
Colle and Salvetti\cite{CAS75}.  The plan being followed is to 
implement these nonlocal potentials for calculations within atomic 
cells, then to use energy-linearized variational multiple
scattering theory to match such local calculations together in
large molecules and solids\cite{NES03}.  This methodology is outlined
here.  Unrestricted Hartree-Fock (UHF) theory can be implemented for
atoms or atomic cells by modifying any standard local-potential atomic
Slater or Kohn-Sham program, to replace local exchange by the nonlocal
Fock operator.  A similar modification allows inclusion of
the correlation energy functional considered here.  Total and
orbital energies computed with this nonlocal exchange-correlation model
are reported for ground states of atoms He through Ne.

\section{Nonlocal potentials in orbital functional theory}
\par The simplest example of a nonlocal functional derivative
is provided by the Schr\"odinger kinetic energy orbital functional
\begin{eqnarray}
T=\sum_in_i(i|{\hat t}|i).
\end{eqnarray}
The notation here sums over occupied orbitals $\phi_i$ of a 
Slater-determinant reference state $\Phi$, with occupation numbers 
$n_i=1$ for $i\leq N$.  Spin indices and sums are assumed, but 
suppressed in the notation.  The total electronic density of the model 
state $\rho({\bf r})=\sum_in_i\rho_i({\bf r})$ is a sum of orbital
densities $\rho_i({\bf r})=\phi^*_i({\bf r})\phi_i({\bf r})$.
The partial (G\^ateaux) density functional derivative
\begin{eqnarray}
\frac{\delta T}{n_i\delta\rho_i}=v_{ti}({\bf r})=
   \frac{\phi^*_i{\hat t}\phi_i}{\phi^*_i\phi_i}.
\end{eqnarray}
is determined as an orbital-indexed local potential  
function\cite{NES02a,NES03a} by the orbital functional derivative
\begin{eqnarray}
\frac{\delta T}{n_i\delta\phi^*_i}={\hat t}\phi_i.
\end{eqnarray}
Applied to the functional $V=\sum_in_i(i|v|i)$, for an external local
potential $v({\bf r})$, this analysis determines the G\^ateaux
derivative $v_{vi}({\bf r})=v({\bf r})$.  Since this is independent of 
the orbital index, it reduces to a Fr\'echet derivative, a 
multiplicative local potential.  The implied generalized Thomas-Fermi
equations for orbital
densities $i\leq N$ of noninteracting electrons are\cite{NES02a}
\begin{eqnarray}
v_{ti}({\bf r})=\epsilon_i-v({\bf r}).
\end{eqnarray}
The Lagrange multipliers $\epsilon_i$ are to be determined such that 
$\int d^3{\bf r}\rho_i=1$ for each orbital density.  These equations 
are operationally equivalent to the noninteracting Kohn-Sham or OEL 
equations\cite{NES02a}
\begin{eqnarray}
{\hat t}\phi_i=\{\epsilon_i-v({\bf r})\}\phi_i.
\end{eqnarray}
\par Two-electron functionals are defined by $U=E_h+E_x$, where 
\begin{eqnarray}\label{Uops}
E_h= \half\sum_{i,j}n_in_j(ij|u|ij);\;
E_x=-\half\sum_{i,j}n_in_j(ij|u|ji),
\end{eqnarray}
and $u=1/r_{12}$.
The Hartree functional $E_h$ is an integral of an explicit function of
total density, which determines the Coulomb potential $v_h({\bf r})$ as
a Fr\'echet derivative.  Similarly, if $E_{xc}$ is defined as in the
LDA, the present analysis determines a Fr\'echet functional derivative, 
and verifies the LDA Kohn-Sham equations.
The orbital functional derivative of the exchange functional $E_x$
defines the Fock exchange operator ${\hat v}_x$ such that
${\hat v}_x\phi_i=-\sum_jn_j(j|u|i)\phi_j$.  The functional derivative 
of $U=E_h+E_x$ defines ${\hat u}=v_h({\bf r})+{\hat v}_x$.  Explicitly,
\begin{eqnarray}
\frac{\delta U}{n_i\delta\phi^*_i}={\hat u}\phi_i=
 \sum_jn_j(j|{\bar u}|j)\phi_i,
\end{eqnarray}
where ${\bar u}=u(1-{\cal P}_{12})$, and ${\cal P}$ is the exchange 
operator for equal spin indices.  
\par Given $(H-E)\Psi=0$ for an N-electron eigenstate and any rule
$\Psi\to\Phi$ that determines a model state $\Phi$, 
unsymmetric normalization $(\Phi|\Psi)=(\Phi|\Phi)=1$ implies 
$E=(\Phi|H|\Psi)=E_0+E_c$.  Here $E_0=(\Phi|H|\Phi)=T+U+V$ is an 
explicit orbital functional, and $E_c=(\Phi|H|\Psi-\Phi)$ defines
the correlation energy.  If ${\cal Q}=I-\Phi\Phi^{\dagger}$, 
$E_c=(\Phi|H|\Psi-\Phi)=(\Phi|H|{\cal Q}\Psi)$.
This implies an exact but implicit orbital functional\cite{NES01a}
\begin{eqnarray}\label{Ecexp}
E_c=-(\Phi|H[{\cal Q}(H-E_0-E_c-i\eta){\cal Q}]^{-1}H|\Phi),
\end{eqnarray}
for $\eta\to0+$.  In practice, some parametrized approximate $E_c$
must be used, defining a correlation potential operator such that 
\begin{eqnarray}
\frac{\delta E_c}{n_i\delta\phi^*_i}={\hat v}_c\phi_i.
\end{eqnarray}
This defines an indexed local potential, the G\^ateaux derivative
\begin{eqnarray}
\frac{\delta E_c}{n_i\delta\rho_i}=v_{ci}({\bf r})=
   \frac{\phi^*_i{\hat v}_c\phi_i}{\phi^*_i\phi_i}.
\end{eqnarray}
Defining universal functional $F=E-V=T+U+E_c$, and an operator 
${\cal F}={\hat t}+{\hat u}+{\hat v}_c$, 
the general OEL equations are
\begin{eqnarray}
\frac{\delta F}{n_i\delta\phi^*_i}=
{\cal F}\phi_i=\{\epsilon_i-v({\bf r})\}\phi_i,
\end{eqnarray}
reducing in the LDA to Kohn-Sham equations.  The OEL equations 
imply generalized Thomas-Fermi equations\cite{NES02a}, 
\begin{eqnarray}
\frac{\phi^*_i{\cal F}\phi_i}{\phi^*_i\phi_i}=v_{fi}({\bf r})=
\epsilon_i-v({\bf r}).
\end{eqnarray}
If ${\cal F}$ is hermitian, the indexed potential $v_{fi}$ is the
G\^ateaux derivative $\delta F/n_i\delta\rho_i$.  There is no 
implication in general that $E_x+E_c$ defines a Fr\'echet derivative.

\section{Plan for large molecules and solids}
The general case of nonlocal potentials determined explicitly by the
idempotent Dirac density matrix
\begin{eqnarray}
{\hat\rho}(1,2)=\sum_i\phi_i({\bf r}_1)n_i\phi^*_i({\bf r}_2)
\end{eqnarray}
is considered.  In full-potential multiple scattering theory 
(MST)\cite{GAB00}, local basis functions are constructed by integrating
the Schr\"odinger or semirelativistic equation for specified orbital
angular momentum $\ell$ and energy $\epsilon$ within the enclosing
sphere ($r=r_1$) of each atomic cell.  Following ideas of canonical
energy-band theory\cite{AND75,AAJ77}, reviewed by 
Skriver\cite{SKR84}, the energy-dependent radial wave function 
$u_{\ell}(\epsilon;r)$ is characterized by its logarithmic derivative 
$D_{\ell}(\epsilon)=r_Su'_{\ell}(r_S)/u_{\ell}(r_S)$,
evaluated at the radius of a sphere whose volume equals that of a
polyhedral atomic cell.  As discussed originally by Wigner
and Seitz, this cellular wave function can continue smoothly across a
cell interface if $D_{\ell}(\epsilon)$ is negative, implying that
$D_{\ell}(\epsilon_B)=0$ and $D_{\ell}(\epsilon_A)=-\infty$ define
the lower and upper energy limits of an energy band.  The band center
is estimated to occur at $D_{\ell}(\epsilon_C)=-\ell-1$.
An equivalent parameter is 
$p_{\ell}(\epsilon)=\frac{D_{\ell}(\epsilon)+\ell+1}
 {D_{\ell}(\epsilon)-\ell}$,
which varies nearly linearly over the width of a band.  It is found
that orbital wave functions are well-approximated by linear 
interpolation over energies $\epsilon_B\leq\epsilon\leq\epsilon_A$.
\par This behavior indicates that an initial self-consistent calculation
should be carried out for spherically averaged potentials, subject to
the orbital boundary condition $D_{\ell}(\epsilon)=-\ell-1$ at $r_S$, in
each inequivalent atomic cell.  Because they satisfy a fixed boundary
condition, the set of eigenfunctions $u_{\ell}(\epsilon_C;r)$ can be
extended to completeness within $r_S$, defining a local basis set for
the linearized variational cellular method (LVCM)\cite{NES03}.  The
computed nonlocal potential must be extended out to the enclosing
sphere $r=r_1$, using the density  matrix ${\hat\rho}$ constructed from
the self-consistent cell orbitals, normalized to unity within $r_S$.  
Orbital functions should be computed, using this fixed potential, for
energies $\epsilon_C$. Energy-derivative functions
${\dot u}_{\ell}(\epsilon_C;r)$ must be computed as a basis
for expanding the LVCM global matching function\cite{NES03,SKR84}.
\par At the UHF level of calculation, in each self-consistent iteration
an average local exchange potential (Slater exchange) is defined by
\begin{eqnarray}
v_x({\bf r})=
 \sum_in_i v_{xi}({\bf r})\rho_i({\bf r})/\sum_in_i\rho_i({\bf r}),
\end{eqnarray}
an exact formula if $v_{xi}$ reduces to a Fr\'echet derivative.  This
approximation is corrected by incremental inhomogeneous terms
\begin{eqnarray}
\{{\hat v}_x-v_x\}\phi_i
\end{eqnarray}
evaluated from the previous iteration.  This procedure converts a 
local-potential algorithm into UHF, and is valid for the indexed 
correlation potential considered below.
\section{Short-range correlation energy}
Electronic correlation energy arises from two quite different sources.
At distances larger than an atomic radius, multipolar response
produces correlation effects evident in polarization potentials and
dispersion forces. Long-range correlation, not considered here, requires
computation of first-order multipole response pseudostates for each 
basis function $u_{\ell}(\epsilon_C;r)$\cite{NES00}.
A different approach is required for the short-range 
correlation due to the singularity of the interelectronic Coulomb
potential $u=1/r_{12}$.  Expansion in $r_{12}$ about such a singularity
shows that an N-electron wave function must have specific cusp 
behavior\cite{KAT57,BIN67} in order to cancel the singularity.
The wave function must vary as $1+\half r_{12}+\cdots$ . 
Colle and Salvetti (CS)\cite{CAS75} impose this cusp condition
through a symmetrical factor 
\begin{eqnarray}
\Pi_{i<j}[1-\xi({\bf r}_i,{\bf r}_j)],
\end{eqnarray}
multiplying an antisymmetric model wave function.  Using coordinates
${\bf q}={\bf r}_i-{\bf r}_j$ and ${\bf r}=\half({\bf r}_i+{\bf r}_j)$
the CS Ansatz is
\begin{eqnarray}
\xi({\bf r},{\bf q})=
 \exp(-\beta^2q^2)[1-\Gamma({\bf r})(1+\half q)].
\end{eqnarray}
This ensures the limiting forms
\begin{eqnarray}
(1-\xi)|_{q\to 0}&=&\Gamma({\bf r})(1+\half q+\cdots) \nonumber\\
\xi|_{q\to\infty}&=&0.
\end{eqnarray}
\par
In adapting this Ansatz to orbital functional theory
(OFT)\cite{NES01,NES01a},
it is desirable to retain the unsymmetric normalization
condition $(\Phi|\Psi)=(\Phi|\Phi)=1$, where $\Psi$ is the correlated 
state, and $\Phi$ is a reference state Slater determinant.  Then the
correlation energy is $E_c=(\Phi|H|\Psi-\Phi)=\sum_{i<j}n_in_jE_{ij}$, 
a sum of electron-pair correlation energies.
This suggests parametrization for each pair of
occupied orbital functions $\phi_i,\phi_j$, such that 
\begin{eqnarray}\label{Eij}
E_{ij}=-(ij|{\bar u}\xi_{ij}(q)|ij). 
\end{eqnarray}
A parametrized form similar to CS is 
\begin{eqnarray}
\xi_{ij}(q)=\exp(-\beta^2_{ij}q^2)[1-\gamma_{ij}(1+\half q)].
\end{eqnarray}
Parameter $\gamma_{ij}$ is determined by the normalization condition
$(\Phi|\Psi-\Phi)=0$, or $(ij|\xi_{ij}|ij)=0$ for each pair: 
\begin{eqnarray} 
\gamma_{ij}=\frac{(ij|\exp(-\beta^2_{ij}q^2)|ij)}
 {(ij|\exp(-\beta^2_{ij}q^2)(1+\half q)|ij)}.
\end{eqnarray}
The free parameter $\beta_{ij}$ can be chosen to minimize 
Bethe-Goldstone (BG or IEPA) energy\cite{SAO89} for specified $i,j$,
or can be treated as a semiempirical parameter and fitted to known
correlation energies of atoms.  
\par This Ansatz can be incorporated into OFT by using 
Eq.(\ref{Eij}) to define the model correlation energy.  At any stage
of the self-consistency iteration, parameter $\gamma_{ij}$ is determined
by the consistency condition given above, and $\beta_{ij}$ is either
a fixed empirical parameter, or is to be updated by solving a 
2-electron BG equation indexed by orthogonal occupied orbitals $ij$.
The indexed local correlation potential 
(G\^ateaux functional derivative) is\cite{NES03a}
\begin{eqnarray}
v_{ci}({\bf r})=\frac{\phi^*_i({\bf r}){\hat v}_c\phi_i({\bf r})}
 {\phi^*_i({\bf r})\phi_i({\bf r})},
\end{eqnarray}
where
\begin{eqnarray}
{\hat v}_c\phi_i=\frac{\delta E_c}{n_i\delta\phi^*_i}=
 -\sum_jn_j(j|{\bar u}\xi_{ij}|j)\phi_i,
\end{eqnarray}
a direct generalization of the operator ${\hat u}$.  Both $v_h$ and
${\hat v}_x$ are modified by short-range correlation, and 
antisymmetry is built in.  Because terms $j=i$ vanish, this
Ansatz for Coulomb-cusp correlation does not produce self-interaction. 
The indexed local potential $v_{ci}({\bf r})$ is singular at nodes of
$\phi_i({\bf r})$.  As in Hartree-Fock methodology, ${\hat v}_c\phi_i$
should be treated as an inhomogeneous term in numerical solution of the
orbital Euler-Lagrange (OEL) equations.
\par Coupling of electron pairs is significant in standard CI methods
for electronic correlation\cite{SAO89}.  Although such coupling is
inherent in the self-consistency of the proposed mean-field model,
a more accurate extension of the method may be needed.  A possible
procedure is to implement the coupled electron-pair approximation
(CEPA)\cite{MEY73} within each atomic cell, using the modified
CS Ansatz as a closure formula for the CI expansion.
\subsection{Notes on integrals}
The normalized orbital basis functions are of the form
\begin{equation}\label{basis}
\phi_a({\bf r})=N_a\chi_a(r)Y_{\ell_a m_a}(\theta,\phi),
\end{equation}
where $r\chi_a(r)$ is a numerical solution of the
radial OEL equation.  The normalization constants are
$N_a=[\int r^2dr \chi_a^2(r)]^{-\half}$.  Definite and indefinite
integrals are required for two-electron generalized potential functions
$F(q)$, where $q^2=r_1^2+r_2^2-2r_1r_2\cos\theta$.  Integration over
angles follows the standard derivation\cite{CAS35} of Condon and
Shortley for $F(q)=1/q$.  For normalized radial functions,
\begin{eqnarray}
(ab|F|cd)=\sum_k c^k(ac)c^k(db)F^k(ac;db),
\end{eqnarray}
where $c^k$ denotes a Gaunt coefficient\cite{CAS35} and
\begin{eqnarray}
F^k(ac;db)=\int_0^\infty r_1^2dr_1 \int_0^\infty r_2^2dr_2
\nonumber\\
 f^k(r_1,r_2) \chi^*_a(r_1)\chi_c(r_1) \chi_d(r_2)\chi^*_b(r_2).
\end{eqnarray}
The factor $f^k$ is
\begin{eqnarray}
f^k(r_1,r_2)=\frac{2k+1}{2}\int_0^\pi \sin\theta d\theta 
 P_k(\cos\theta) F(q).
\end{eqnarray}
Given $r_1$ and $r_2$, 
$\cos\theta=\frac{r_1^2+r_2^2-q^2}{2r_1r_2}$
and 
$\sin\theta d\theta=\frac{qdq}{r_1r_2}$.  In internal 
coordinates, $f^k(r_1,r_2)$ is an integral of the form 
\begin{eqnarray}
\frac{2k+1}{2r_1r_2}\int_{|r_1-r_2|}^{r_1+r_2}
 P_k(\frac{r_1^2+r_2^2-q^2}{2r_1r_2})qF(q)dq.
\end{eqnarray}
These integrals are needed for $qF(q)$ equal to $1,q,q^2$ times
the factor $\exp(-\beta^2q^2)$, so that the integrand of $f^k$ is
this Gaussian factor times a polynomial in $q$.  It can easily be
verified that $f^k$ reduces to $r_<^k/r_>^{k+1}$ for $\beta\to0$ 
if $qF(q)=1$.  The Legendre polynomial factors of the integrand
are determined by the recurrence formula
\begin{eqnarray}
P_{k+1}(x)=\frac{2k+1}{k+1}xP_k(x)-\frac{k}{k+1}P_{k-1}(x),
\end{eqnarray}
with $P_{-1}(x)=0, P_0(x)=1$.
The elementary integrals required are of the form
$I_n(\beta;\kappa)=\int_0^\kappa e^{-\beta^2q^2}q^ndq$.  By a change
of variables such that $t=\beta^2q^2$, this reduces to 
\begin{eqnarray}
I_n(\beta;\kappa)&=&
 \frac{1}{2\beta^{n+1}}\int_0^{\beta^2\kappa^2}e^{-t}t^{\half(n-1)}dt
\nonumber\\&=&
 \frac{\Gamma(\half(n+1))}{2\beta^{n+1}}
 p(\half(n+1),\beta^2\kappa^2),
\end{eqnarray}
in terms of the incomplete gamma function 
$p(a,x)=1-\Gamma(a,x)/\Gamma(a)$\cite{PTVF92}, Sect.6.2.  $p(a,x)$ can
be computed effficiently using a power series for small $x$ and a
continued fraction for large $x$.  The continued fraction terminates 
if $n$ is odd. 

\section{Calculations on light atoms}
For applications to molecules and solids, using multiple scattering 
theory and an atomic cell model, values of the parameters $\beta_{ij}$ 
can be obtained by calculations on atoms.  Results of 
such calculations, for light atoms He through Ne, are reported
here.  A numerical Dirac-Slater program\cite{DES69} was modified as
described above for UHF (exchange-only) and OFT calculations, the latter
incorporating the modified CS correlation energy functional described
above.  The program was used in its nonrelativistic mode.  Angular
coefficients in the total energy functional were computed such that
effective potentials are spherically averaged, but retain a spin
index.  This equivalence restriction implies that radial orbital
functions with indices $\ell,m_s$ are well-defined.
\par For each atom considered, three sets of self-consistent
calculations were carried out:
for the He-like ion, for the Be-like ion (for
$N\geq4$), and for the neutral atom.  To verify the
computational method, computed UHF energies are compared in 
Table(\ref{tab01}) with established RHF energies\cite{CAR74} and with
total energies including correlation\cite{CLE63}.  The OFT
calculations were used to determine $\beta_{ij}$ parameters such that
the computed total energies agreed with the "experimental" values shown
in Table(\ref{tab01}). Pair-indexed parameters $\beta_{1s1s}$ were
determined for He-like ions, parameters $\beta_{2s2s}$ for Be-like
ions, and parameters $\beta_{2p2p}$ for the neutral atoms.  In each
case, inner parameters were frozen and intershell parameters such
as $\beta_{1s2s}$ were scaled to the geometric mean of the corresponding
intrashell parameters.  Values of the latter that fit total energies of
the ions and atoms considered are listed in Table(\ref{tab02}).  These
parameters are the principal result of the present calculations, 
intended to define parametrized correlation functionals for extended
systems.
\par 
Self-consistent UHF orbital energies are tabulated in Table(\ref{tab03})
and OFT orbital energies in Table(\ref{tab04}).  Because Janak's
theorem\cite{JAN78} is valid in OFT, these energies have a physical 
meaning in the context of a theory in which orbital occupation numbers
are allowed to change continuously and to have fractional values.
They are derivatives of the total energy with respect to infinitesimal
changes of the occupation numbers.  Physical energy 
differences, by implication, correspond to integrals of these
derivatives over finite increments of occupation numbers. 

\section{Conclusions} 
This paper considers an independent-electron model that incorporates
a theoretically motivated Ansatz for correlation energy, expressed as
a parametrized orbital functional.  Computed results
extend exchange-only theory (UHF) to a formalism parametrized by
exact atomic ground-state energies.  Parameters are obtained that make
it possible to apply this formalism to calculations of the electronic
structure of molecules and solids. 

\section*{Acknowledgments}
This work was initiated at the University of Connecticut, supported
by a grant from the University of Connecticut Research Foundation.
The author is grateful to H. H. Michels for discussions
of the new methodology proposed here, and to Prof. Michels and
Prof. Wm. C. Stwalley for encouraging and implementing a visiting
appointment at Storrs. 
%\vfill\eject
% **************** VM screen width ************************************%

% **************** VM screen width ************************************%
\vfill\eject
\begin{table}
\caption{Total energies in Hartree units}
\label{tab01}
\begin{tabular}{lrrr}
Atom&RHF&UHF&exp\\
\tableline
He&-2.8617&-2.8617&-2.9038\\ 
Li&-7.4327&-7.4328&-7.4780\\
Be&-14.5730&-14.5730&-14.6674\\
B &-24.5291&-24.5293&-24.6541\\
C &-37.6886&-37.6900&-37.8466\\
N &-54.4009&-54.4045&-54.5890\\
O &-74.8094&-74.8136&-75.0674\\
F &-99.4093&-99.4108&-99.7333\\
Ne&-128.5470&-128.5470&-128.9400\\
\end{tabular}
\end{table}
%%%%
\begin{table}
\caption{Orbital $\beta$ parameters}
\label{tab02}
\begin{tabular}{lccc}
Atom&1s&2s&2p\\
\tableline
He&0.83455&&\\ 
Li&1.42494&1.16853&\\
Be&2.01812&0.52267&\\
B &2.61268&0.71557&0.82167\\
C &3.20278&0.89701&1.08430\\
N &3.79877&1.05841&1.39511\\
O &4.39600&1.20737&1.55724\\
F &4.99306&1.34754&1.77434\\
Ne&5.57807&1.47974&2.00125\\
\end{tabular}
\end{table}
%%%%
%\onecolumn
\begin{table}
\caption{UHF orbital energies (Hartree units)}
\label{tab03}
\begin{tabular}{lcccccc}
Atom&
$1s_\alpha$&$1s_\beta$&$2s_\alpha$&$2s_\beta$&$2p_\alpha$&$2p_\beta$\\
\tableline
He&-0.91796&-0.91796&\\ 
Li&-2.48668&-2.46870&-0.19637&\\
Be&-4.73267&-4.73267&-0.30927&-0.30927&\\
B &-7.70036&-7.68527&-0.54022&-0.44175&-0.31671&\\
C &-11.34480&-11.29972&-0.82450&-0.57914&-0.43818&\\
N &-15.67067&-15.58098&-1.16297&-0.72580&-0.57092&\\
O &-20.70635&-20.62807&-1.41447&-1.07069&-0.67505&-0.52144\\
F &-26.40567&-26.35786&-1.66855&-1.47217&-0.76666&-0.67975\\
Ne&-32.77237&-32.77237&-1.93040&-1.93040&-0.85041&-0.85041\\
\end{tabular}
\end{table}
\begin{table}
\caption{OFT orbital energies (Hartree units)}
\label{tab04}
\begin{tabular}{lcccccc}
Atom&
$1s_\alpha$&$1s_\beta$&$2s_\alpha$&$2s_\beta$&$2p_\alpha$&$2p_\beta$\\
\tableline
He&-0.94667&-0.94667&\\ 
Li&-2.52396&-2.50794&-0.19546&\\
Be&-4.81661&-4.81661&-0.31196&-0.31106&\\
B &-7.80392&-7.79277&-0.53861&-0.45076&-0.32004&\\
C &-11.46803&-11.43171&-0.81782&-0.59356&-0.44126&\\
N &-15.80892&-15.73252&-1.15024&-0.74332&-0.57450&\\
O &-20.86288&-20.79561&-1.39967&-1.08080&-0.67795&-0.54422\\
F &-26.58113&-26.53950&-1.65541&-1.47252&-0.77354&-0.69690\\
Ne&-32.96943&-32.96943&-1.92025&-1.92025&-0.86284&-0.86284\\
\end{tabular}
\end{table}
\end{document}